\documentstyle[axodraw]{article}
\newcommand{\Tr}{\mathop{\rm Tr}\nolimits}
\newcommand{\para}{_\parallel}
\newcommand{\pr}{_\perp}

\title{\centerline{\small SINP/TNP/02-28, IMSc/2002/07/21 \hfill}
\bf The Axialvector-Vector Amplitude and Neutrino Effective Charge in
a Magnetized Medium}

\bigskip
\author{ Kaushik Bhattacharya$^1$, Avijit K.~Ganguly$^2$\thanks{e-mail
addresses: kaushikb@theory.saha.ernet.in, avijit@imsc.ernet.in }\\
\normalsize
$1)$  Saha Institute of Nuclear Physics, 1/AF, Bidhan-Nagar, Calcutta
700064, India\\
\normalsize
$2)$ Institute of Mathematical Sciences, Chennai, India
}
\begin{document}

\maketitle
\begin{abstract}
To one loop the effective neutrino photon interaction takes place
through the vector-vector type and the axialvector-vector type
amplitude. In this work we explicitly write down the form of the
axialvector-vector amplitude to all orders in the external magnetic
field in a medium. We then infer upon its zero external momentum limit
which contributes to the effective charge of the neutrinos inside a
magnetized medium. We further show its gauge invariance properties.
\end{abstract}
\bigskip

\section{Introduction}
\label {intr}
Neutrino mediated processes are of great importance in cosmology and
astrophysics{\cite{raff,weinberg}}. Various interesting possibilities
involving neutrinos has been looked  in the context of cosmology
e.g, large scale structure formation in the universe, to name one of
the few~\cite{palone}. In this note we would consider an induced
effect on the neutrinos by a magnetized medium, using quantum field
theoretic techniques. It is usually conjectured, taking into account
the conservation of surface magnetic field of a proto-neutron star,
that during a supernova collapse the magnetic field strength in some
regions inside the nascent star can reach upto ${\mathcal B}\sim \frac{
m^2}{e} $ or more. Here $m$ denotes the mass of electron. This
conjecture makes it worthwhile to investigate the role of magnetic
field in effective neutrino photon vertex.

Neutrinos do not couple to photons at the tree level in the standard
model of particle physics, and this coupling can only take place at a
loop level, mediated by the fermions and gauge bosons. This coupling
can give birth to scattering process like $\gamma \gamma \to \nu
\nu$. The cross-section of neutrino-photon scattering is highly
suppressed in the standard model due to Yang's theorem\cite{yang},
which makes the scattering cross-section vanish to orders of the Fermi
coupling $G_F$. In presence of a magnetic field, neutrino photon
scattering can occur and to orders of $G_F$ the cross-section has been
calculated \cite{Shaisultanov:1997bc}. There can be neutrino processes
in a medium or a magnetic field or both which involves only one photon
as $\nu \to \nu \gamma$ and $\gamma \to \nu \bar{\nu}$. In vacuum
these reactions are restrained kinematically. Only in presence of a
medium or a magnetic field or both can all the particles be onshell as
there the dispersion relation of the photon changes, giving the much
required phase space for the reactions.  Intuitively when a neutrino
moves inside a thermal medium composed of electrons and positrons,
they interact with these background particles. The background
electrons and positrons themselves have interaction with the
electromagnetic fields, and this fact gives rise to an effective
coupling of the neutrinos to the photons. Under these circumstances
the neutrinos may acquire an ``effective electric charge'' through
which they interact with the ambient plasma.

The effective charge of the neutrino in a medium has been calculated
previously by many authors ~\cite{pal1,orae,alth}.  All of these works
were concentrated on the vector-vector part of the interaction. In
this paper we concentrate upon the effective neutrino photon
interaction vertex coming from the axialvector-vector part $\Pi^5_{\mu
\nu}$ of the interaction. Some work on the axialvector-vector part in
a time independent background electro-magnetic field has been done
previously\cite{Shaisultanov:2000mg, Schubert:2000kf} where the in
some cases the authors were able to obtain a gauge invariant
expression of the axialvector-vector contribution in the neutrino
photon effective vertex.  We are interested in the zero momentum limit
of the axialvector-vector amplitude, as it contributes to the
effective charge of the neutrinos inside a magnetized plasma. We
discuss the physical situations where the axialvector-vector amplitude
arises, then show how it affects the physical processes.

The plan of the paper is as follows.  We start with Section \ref{form}
that deals with the formalism through which the physical importance of
$\Pi^5_{\mu \nu}(k)$ is appreciated. In Section \ref{gd} general form
factor analysis of the second rank tensor on the basis of
symmetry arguments is provided.  In Section \ref{capt} we show the
fermion propagator in a magnetized medium, and using it explicitly
write down $\Pi^5_{\mu \nu}(k)$ in the rest frame of the medium. In
Section \ref{efftcharge} we calculate the effective electric charge
from the expression of the axialvector-vector amplitude. In Section
\ref{concl} we discuss our results and conclude by touching upon the
physical relevance of our work. A formal proof of the gauge invariance
of $\Pi^5_{\mu\nu}$ is attached in the appendix.
\section{Formalism}
\label{form}
%
\begin{figure}
\begin{center}
\begin{picture}(150,40)(0,-35)
\Photon(40,0)(0,0){2}{4}
\Text(20,5)[b]{$k\leftarrow$}
\Text(75,30)[b]{$p$}
\Text(125,-17)[b]{$q$}
\Text(125,15)[b]{$q'$}
\Text(47,0)[]{$\nu$}
\Text(102,0)[]{$\mu$}
\Text(75,-30)[t]{$p+k\equiv p'$}
\ArrowLine(110,0)(160,30)
\ArrowLine(160,-30)(110,0)
\SetWidth{1.2}
\Oval(75,0)(25,35)(0)
\ArrowLine(74,25)(76,25)
\ArrowLine(76,-25)(74,-25)
\end{picture}
\end{center} 
\caption[]{One-loop diagram for the effective electromagnetic vertex
of the neutrino in the limit of infinitely heavy W and Z
masses.}\label{f:cher}
\end{figure}
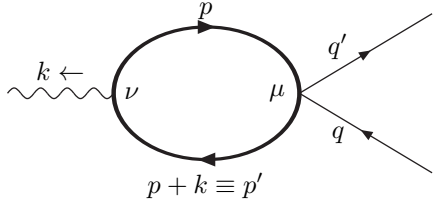
In this work we consider the background temperature and neutrino
momenta that are small compared to the masses of the W and Z
bosons. We can, therefore, neglect the momentum dependence in the W
and Z propagators, which is justified if we are performing a
calculation to the leading order in the Fermi constant, $G_F$. In this
limit four-fermion interaction is given by the following effective
Lagrangian:
\begin{eqnarray}
{\cal L}_{\rm eff} = -\frac{1}{\sqrt{2}} G_F {\overline \nu}
\gamma^{\mu} (1 - \gamma_5) \nu \,\,{\overline l} \gamma_{\mu} (g_{\rm
V} + g_{\rm A} \gamma_5) l \,,
\label{effl}
\end{eqnarray}
where, $\nu$ and $l$ are the neutrino and the corresponding lepton
fields. For electron neutrinos,
\begin{eqnarray}
g_{\rm V} &=& \frac{1}{2} + 2 \sin^2 \theta_{\rm W},\nonumber \\
g_{\rm A} &=& - \frac{1}{2}\nonumber
\end{eqnarray}
where the first terms in $g_{\rm V}$ and $g_{\rm A}$ are the
contributions from the W exchange diagram and the second one from the
Z exchange diagram. $\theta_{\rm W}$ is the Weinberg angle. For
muon and tau neutrinos 
\begin{eqnarray}
g_{\rm V} &=& -\frac{1}{2} + 2 \sin^2 \theta_{\rm W},\nonumber \\
g_{\rm A} &=&  \frac{1}{2}\,.\nonumber
\end{eqnarray}

With this interaction Lagrangian the $\nu \nu \gamma$ vertex, as
shown in fig.~[\ref{f:cher}], can be written in terms of two
tensors. The vector-vector amplitude and the axialvector-vector
amplitude.The vector-vector amplitude term $\Pi_{\mu \nu}(k)$ is defined
as
\begin{eqnarray}
i\Pi_{\mu \nu}(k)=(-i e)^2(-1) \int {{d^4 p}\over {(2\pi)^4}}\Tr \left
[ \gamma_\mu iS(p) \gamma_\nu iS(p')\right]\,.
\label{npimunu}
\end{eqnarray}
Here and henceforward $p'= p + k$. The above equation looks exactly
like the photon polarization tensor, but dosent have the same
interpretation here. The Fynman diagram associated with the neutrino
photon interaction to one loop is depicted in fig.[\ref{f:cher}]. The
axial-vector--vector amplitude $\Pi^5_{\mu \nu}(k)$ is defined as
\begin{eqnarray}
i\Pi^5_{\mu \nu}(k)=(-i e)^2(-1) \int {{d^4 p}\over {(2\pi)^4}}\Tr \left
[ \gamma_\mu \gamma_5 iS(p) \gamma_\nu iS(p')\right].
\label{npimunu5}
\end{eqnarray}
Both tensors are obtained by calculating the Feynman diagram given in
fig.[\ref{f:cher}].

While discussing about $\Pi^5_{\mu \nu}(k)$ it should be remembered that for
the electromagnetic vertex, we  have the current conservation relation,
\begin{eqnarray}
k^{\nu} \Pi^5_{\mu \nu}(k) = 0
\label{ngi}
\end{eqnarray}
which is the gauge invariance condition.

In order to calculate the effective charge of the neutrinos inside a
medium, we have to calculate $\Pi^5_{\mu \nu}(k)$. The formalism so
discussed is a general one and we extend the calculations previously
done based upon this formalism to the case where we have a constant
background magnetic field in addition to a thermal medium.
\section{General form of $\Pi^{5}_{\mu\nu}(k)$ in various cases}
\label{gd}
\subsection{The vacuum case}
We start this section with a discussion on the possible tensor
structure and form factor analysis of $\Pi^{5}_{\mu\nu}(k)$, based on
the symmetry of the interaction. To begin with we note that,
$\Pi^{5}_{\mu\nu}(k)$ in vacuum should vanish.  This follows from the
following arguments. In vacuum the available vectors and tensors at
hand are the following,
\begin{eqnarray}
k_{\mu},~g_{\mu\nu} \mbox{~and~} \epsilon_{\mu\nu\lambda\sigma}.
\label{vt}
\end{eqnarray}

The two point axialvector-vector amplitude $\Pi^{5}_{\mu\nu}$ can be
expanded in a basis, constructed out of the above tensors. Given the
parity structure of the theory the only relevant tensor at hand is
$\epsilon_{\mu\nu\lambda\sigma}k_{\lambda}k_{\sigma}$ which is zero.
\subsection{In medium}
On the other hand, in a medium in absence of any magnetic field, we
have an additional vector $u^{\mu}$, i.e the velocity of the centre of
mass of the medium. Therefore the axialvector-vector amplitude can be
expanded in terms of form factors along with the new tensors
constructed out of $u^{\mu}$ and the ones we already had in absence of
a medium. A second rank tensor constructed out of them can be
$\varepsilon_{\mu \nu \alpha \beta}u^{\alpha}k^{\beta}$, \cite{pal1}
which would verify the current conservation condition for the two
point function. In a medium an interesting thing happens. In the
tensor $\Pi^5_{\mu \nu}(k)$ one of the tensor indices refers to the
vector type vertex another one to the axialvector type. The
electromagnetic current conservation condition is supposed to hold for
the vector type vertex. But due to the tensor structure of $\Pi^5_{\mu
\nu}(k)$ in a medium, as discussed
\begin{eqnarray}
k^{\mu} \Pi^5_{\mu \nu}(k) = 0 = \Pi^5_{\mu \nu}(k) k^{\nu}
\label{mgi}
\end{eqnarray}
to all orders in the Fermi and electromagnetic coupling. 
\subsection{In a background magnetic field}
As in our case neither {\bf C} or {\bf P} but {\bf CP} is a symmetry,
first we look at the {\bf CP} transformation properties of the
axialvector-vector amplitude. Under a {\bf CP} transformation the
various components of the tensor transforms as,
\begin{eqnarray}
{\rm{CP}}:\,\Pi^5_{0\,0} &\to& \Pi^5_{0\,0}\,,
\label{pi5cp0}\\
{\rm{CP}}:\,\Pi^5_{i\,j} &\to& \Pi^5_{i\,j}\,,
\label{pi5cpij}\\
{\rm{CP}}:\,\Pi^5_{0\,i} &\to& -\Pi^5_{0\,i}.
\label{pi5cp0i}
\end{eqnarray}
The above transformation properties are important because the basis
tensors must also satisfy the same transformation properties. The
functions which multiply the basis tensors to build up
$\Pi^5_{\mu\nu}(k)$, called the form factors, must be even functions of
magnetic field and so their {\bf CP} transformations are trivial. This
fact directly implies the basis tensors must vanish in the ${\mathcal
B} \to 0$ limit, as $\Pi^5_{\mu \nu}(k)$ vanishes in vacuum. In general
$\Pi^5_{\mu \nu}(k)$ will be an odd function of the external field so
that it vanishes in the external field going to zero limit. As a
result the basis tensors must also be odd in the external fields.

In a uniform background magnetic field, the vectors
and tensors at hand are
\begin{eqnarray}
F_{\mu\nu},\qquad {\widetilde{F}_{\mu\nu}},\qquad k^{\mu}_{\para},
\qquad k^{\mu}_{\pr}.   
\end{eqnarray}
Here $F_{\mu\nu}$ is the electromagnetic field strength tensor and
 ${\widetilde F}^{\mu \nu} = \frac{1}{2}{\varepsilon^{\mu \nu \rho
 \sigma}} F_{\rho \sigma}$. Here we stick to the fact that the
 magnetic field is in the $z$ direction, and so
\begin{eqnarray}
F_{1\,2} = - F_{2\,1} =  \mathcal{B}\,.
\end{eqnarray}
with all other components of $F_{\mu\nu}$  zero.   

With this building blocks we can
build four vectors,
\begin{eqnarray}
b^{\mu}_1 &=& (Fk)^{\mu}\,, \label{b1}\\
b^{\mu}_2 &=& (\widetilde{F}k)^{\mu}\,,\label{b2}\\
b^{\mu}_3 &=& k^{\mu}_{\para}\,,\label{b3}\\
b^{\mu}_4 &=& k^{\mu}_{\pr}.
\label{b4}
\end{eqnarray}
The expressions $(Fk)_{\mu}$, and $({\widetilde F}k)_{\mu}$ stands for
\begin{eqnarray}
(Fk)_{\mu}&=& F_{\mu \nu} k^{\nu}\,,\nonumber\\
({\widetilde F}k)_{\mu}&=&{\widetilde F}_{\mu \nu} k^{\nu}.
\end{eqnarray}
Here $k^{\mu}\para = (k^0, 0,0, k^3)$ and $k^{\mu}\pr = (0, k^1, k^2,
0)$, so that $k^{\mu} = k^{\mu}\para + k^{\mu}\pr$. Also in our
convention
\begin{eqnarray}
g_{\mu \nu} = g^{\para}_{\mu \nu} + g^{\pr}_{\mu \nu}\nonumber\,,
\end{eqnarray}
where
\begin{eqnarray}
g^{\para}_{\mu \nu} & = & (1,0,0,-1)\nonumber\\
g^{\pr}_{\mu \nu} & = & (0,-1,-1,0)\,.\nonumber
\end{eqnarray}
For also use,
\begin{eqnarray}
k\para^2 &=& k_0^2 - k_3^2 \nonumber \\
k\pr^2 &=& k_1^2 + k_2^2 \nonumber\,.
\end{eqnarray}

The set of four vectors $b^{\mu}_i$s, $i=1,2,3,4$ are mutually
orthogonal to each other and can serve as the basis vectors to build
up the tensor basis of $\Pi^5_{\mu \nu}(k)$.

Next the {\bf CP} transformation properties of these vectors are
summarized.
\begin{eqnarray}
{\rm{CP}}:\,b^0_1 &\to& b^0_1\,,\\
{\rm{CP}}:\,b^i_1 &\to& b^i_1.
\end{eqnarray}
The other three vectors have similar transformation propertis as
\begin{eqnarray}
{\rm{CP}}:\,b^0_{1,2,3} &\to&   b^0_{1,2,3}\,\\
{\rm{CP}}:\,b^i_{1,2,3} &\to& - b^i_{1,2,3}.
\end{eqnarray}
From Eq.~(\ref{pi5cp0}), Eq.~(\ref{pi5cpij}) and Eq.~(\ref{pi5cp0i})
we can see that a suitable tensor basis can be built up from vectors
$b^{\mu}_i$ where $i = 2,3,4$. The {\bf CP} transformation of the
axialvector-vector amplitude compels us to disregard $b^{\mu}_1$ as a
basis vector.

Now we can list the possible candidates which can serve as basis
tensors of $\Pi^5_{\mu \nu}(k)$. There are nine of them. For later
usage we clearly write them down as,
\begin{eqnarray}
{\rm B}^{\mu \nu}_1 &=& b^{\mu}_2 b^{\nu}_2\nonumber\\
                    &=& (\widetilde{F}k)^{\mu}(\widetilde{F}k)^{\nu}\,,
\label{B1}\\
{\rm B}^{\mu \nu}_2 &=&  b^{\mu}_3 b^{\nu}_3\nonumber\\
                    &=&  k^{\mu}_{\para} k^{\nu}_{\para}\,,
\label{B2}\\
{\rm B}^{\mu \nu}_3 &=& b^{\mu}_4 b^{\nu}_4\nonumber\\ &=&
                    k^{\mu}_{\pr} k^{\nu}_{\pr}\,,
\label{B3}\\
{\rm B}^{\mu \nu}_4 &=& b^{\mu}_2 b^{\nu}_3\nonumber\\ &=&
                    (\widetilde{F}k)^{\mu} k^{\nu}_{\para}\,,
\label{B4}\\
{\rm B}^{\mu \nu}_5 &=& b^{\mu}_3 b^{\nu}_2\nonumber\\ &=&
                    (\widetilde{F}k)^{\nu} k^{\mu}_{\para}\,,
\label{B5}\\
{\rm B}^{\mu \nu}_6 &=& b^{\mu}_2 b^{\nu}_4\nonumber\\ &=&
                    (\widetilde{F}k)^{\mu} k^{\nu}_{\pr}\,,
\label{B6}\\
{\rm B}^{\mu \nu}_7 &=& b^{\mu}_4 b^{\nu}_2\nonumber\\ &=&
                    (\widetilde{F}k)^{\nu} k^{\mu}_{\pr}\,,
\label{B7}\\
{\rm B}^{\mu \nu}_8 &=& b^{\mu}_3 b^{\nu}_4\nonumber\\ &=&
                    k^{\mu}_{\para} k^{\nu}_{\pr}\,,
\label{B8}\\
{\rm B}^{\mu \nu}_9 &=& b^{\mu}_4 b^{\nu}_3\nonumber\\ &=&
                    k^{\nu}_{\para} k^{\mu}_{\pr}\,.
\label{B9}           
\end{eqnarray}
This basis gives nine second rank mutually orthogonal tensors. Any
second rank tensor containing higher field dependence can be
represented by suitable linear combinations of these tensors.

Out of these nine basis tensors some are useless. To explain the point
we focus our attention on ${\rm B}^{\mu \nu}_2$, ${\rm B}^{\mu
\nu}_3$, ${\rm B}^{\mu \nu}_8$ and ${\rm B}^{\mu \nu}_9$. All of the
four mentioned ones do not vanish in the ${\mathcal B} \to 0$ limit
and so they are redundant. Also ${\rm B}^{\mu \nu}_1$ is even in the
external fields, and as discussed previously is not a suitable
candidate for the basis of $\Pi^5_{\mu \nu}(k)$.

Only four basis tensors qualify succesfully as the building blocks of
the axilavector-vector amplitude. They are ${\rm B}^{\mu \nu}_4$,
${\rm B}^{\mu \nu}_5$, ${\rm B}^{\mu \nu}_6$ and ${\rm B}^{\mu
\nu}_7$. The result as given in the papers by Hari Dass and Raffelt
verifies this choice{\cite{hari,iorf}}\footnote{However
the metric used by the authors in refernces mentioned is different
from that of ours, and in their calculation the $\mu$ vertex is the
vector type vertex.}.
\begin{eqnarray}
& &\Pi^{5}_{\mu\nu}(k) = \frac{e^3}{(4\pi)^2 m^2}\left[-C\para
k_{\nu\para} ({\widetilde F}k)_{\mu}\right.\nonumber\\
& &\left.\hskip 1cm +\,\, C\pr\left\{k_{\nu\pr}
(k{\widetilde F})_{\mu} + k_{\mu\pr}(k{\widetilde F})_{\nu} -
k^2\pr{\widetilde F}_{\mu \nu}  \right\}\right]
\label{harid}
\end{eqnarray}
In the results we find ${\widetilde F}^{\mu
\nu}$ which we have not listed in our tensor basis, but that is not a
fault because it can be made up from the basis supplied. ${\widetilde
F}^{\mu \nu}$ can be written
\begin{eqnarray}
{\widetilde F}^{\mu \nu} = \frac{1}{k^2_{\para}}\left[(\widetilde{F}k)^{\mu}
k^{\nu}_{\para} - (\widetilde{F}k)^{\nu} k^{\mu}_{\para} \right]\,.   
\end{eqnarray}
So to build up the tensorial basis of the axialvector-vector
amplitude, the number of independent tensors recquired are four.

The four tensors at hand are still not suitable to be the basis
tensors of the axialvector-vector amplitude. As all of them are not
transverse to $k^{\mu}$, which is a recquirement from electromagnetic
current conservation. Furthermore we have to make linear combinations of
these tensors which will ultimately give two tensors orthogonal to
each other and to $k^{\mu}$, which will serve as the right tensor
basis of $\Pi^5_{\mu \nu}(k)$.
\subsection{In a magnetized medium}
In presence of a magnetized medium the situation complicates. In this
analysis we are not going into an indepth study of the tensorial basis
as is done in the case where there was no medium. Here we outline the
strategy in which we can build up the tensorial basis, which in
reality is similar to the previous case but contains more building
blocks.  To start with we again emphasize on the {\bf CP}
transformation properties of the axialvector-vector amplitude. Unlike
the vacuum case now the theory may not be {\bf CP} invariant. This can
arise if the background does not respect {\bf CP}. We will discuss
here only those cases where the background does not break {\bf
CP}. Moreover now the form factors can be functions of odd powers of
the magnetic field, as now new scalars as $(Fk)u$ and $({\widetilde
F}k)u$ are also available. These scalars change sign under {\bf CP}
transformation. Some of the form factors containing odd order of the
external fields may be accompanied by equal powers of chemical
potential of the background charged fermions, and they will not change
sign. So in a magnetized medium there can be basis tensors with
different {\bf CP} transformation properties as the form factors which
multiply them can also have different transformation properties.

In presence of a medium, we can have two sets of orthogonal
vectors. The first set is as supplied in Eq.~(\ref{b1}),
Eq.~(\ref{b2}), Eq.~(\ref{b3}) and Eq.~(\ref{b4}). They are all
included now. $b^{\mu}_1$ is not excluded as in vacuum because the
{\bf CP} transformation property of the basis tensors have
changed. The other set of orthogonal vectors useful in a medium, are
\begin{eqnarray}
b'^{\mu}_1 &=& (\widetilde{F}u)^{\mu}\,,\label{bp2}\\
b'^{\mu}_2 &=& u^{\mu}_{\para}\,,\label{bp3}
\label{bp4}
\end{eqnarray}
In listing the above vectors we have omitted two vectors. One is $(F
u)^{\mu}$ and $u^{\mu}\pr$. The reason they are omitted is that,
ultimately we are interested in the rest frame of the medium. In the
medium rest frame there is no electric field. Also in medium rest
frame no contribution will come with $u^{\mu}\pr$. 

This above set of vectors has similar {\bf CP} transformation
properties with those of $b^{\mu}_2$, and $b^{\mu}_3$. But the two
sets of vectors are not linearly independent and as such cannot serve
as basis vectors to build up the tensorial basis of $\Pi^5_{\mu
\nu}$. Only a linear combination of them can make a orthogonal vector
basis. Now we list the set of orthogonal basis vectors which can be
made from the two set of vectors, they are
\begin{eqnarray}
b''^{\mu}_1 &=& (Fk)^{\mu}\,, \label{bpp1}\\
b''^{\mu}_2 &=& ({\widetilde F}u)^{\mu} + ({\widetilde F}k)^{\mu}\,,
\label{bpp2}\\ 
b''^{\mu}_3 &=& k^{\mu}_{\pr}\,,\label{bpp3}\\
b''^{\mu}_4 &=& k^{\mu}_{\para} + u^{\mu}_{\para}\,.
\label{bpp4}
\end{eqnarray}
In a magnetized medium we have these four basis vectors which serves
as the building blocks of the axialvector-vector amplitude. The basis
tensors in this case will be the direct product of these basis
vectors. There will be sixteen of them but all of them will not be
useful.

As was the case in vacuum, all the sixteen basis tensors here are not
useful because we have the electromagnetic current conservation
condition. This constraint will reduce the number of basis
tensors. The axilvector-vector amplitude is not orthogonal to
$k^{\mu}$ in our case, as the $\mu$ vertex is the axial vector vertex,
but then still
\begin{eqnarray}
k^{\mu} \Pi^5_{\mu \nu} = {\cal C}_{\nu}
\end{eqnarray}
for some ${\cal C}_{\nu}$ which depends on the mass of the looping
fermions. This condition also restrains the number of basis tensors of
the axilvector-vector amplitude in a magnetized medium.
The exact calculations of the number of useful elements as
basis now goes in the same way we had done it in absence of a medium. 
\section{One  loop calculation of the  axialvector-vector amplitude}
\label{capt}
Since we investigate the case with a uniform background magnetic
field, without any loss of generality it can be taken to be in the
$z$-direction. We denote the magnitude of this field by ${\mathcal
B}$, which can be incorporated in various gauges with $A_0 = 0$ and
the other components of $A$ being time independent. Ignoring first the
presence of the medium, the electron propagator in such a field can be
written down following Schwinger's approach~\cite{schw,tsai,ditt}:
\begin{eqnarray}
i S_B^V(p) = \int_0^\infty ds \, e^{\Phi(p,s)} \, G(p,s) \,,
\label{SV}
\end{eqnarray}
where $\Phi$ and $G$ are as given below 
\begin{eqnarray}
\Phi(p,s) &\equiv& 
          is \left( p_\parallel^2 - {\tan (e{\mathcal B}s) \over e{\mathcal B}s} \, p\pr^2 - m^2 \right) - \epsilon |s| \,,
\label{Phi} \\
G(p,s) &\equiv& {e^{ie{\mathcal B}s\sigma\!_z} \over \cos(e{\mathcal B}s)} 
       \, \left( \rlap/p_\parallel + \frac{e^{-ie{\mathcal B}s\sigma_z}}
{\cos(e{\mathcal B}s)}\rlap/ p\pr + m \right) \nonumber \\ 
       &=& \Big[ \big( 1 + i\sigma_z \tan (e{\mathcal B}s) \big)
(\rlap/p_\parallel + m ) +\sec^2(e{\mathcal B}s) \rlap/ p\pr \Big] \,, 
\label{C}
\end{eqnarray}
where
\begin{eqnarray}
\sigma_z = i\gamma_1 \gamma_2 = - \gamma_0 \gamma_3 \gamma_5 \,,
\label{sigz}
\end{eqnarray}
and we have used,
\begin{eqnarray}
e^{ie{\mathcal B}s\sigma_z} = \cos( e{\mathcal B}s) + i\sigma_z \sin(e{\mathcal B}s) \,.
\end{eqnarray}
To make the expressions transparent we specify our convention in the
following way,
\begin{eqnarray}
\rlap/ p_\parallel &=& \gamma_0 p^0 + \gamma_3 p^3 \nonumber \\                      
\rlap/p\pr &=& \gamma_1 p^1 + \gamma_2 p^2 \,.\nonumber \\
\end{eqnarray}
Of course in the range of integration indicated in Eq.~(\ref{SV}) $s$
is never negative and hence $|s|$ equals $s$.
In the presence of a background medium, the above
propagator is now modified to~\cite{elmf}:
\begin{eqnarray}
iS(p) = iS_B^V(p) + S_B^\eta(p) \,,
\label{fullprop}
\end{eqnarray}
where
\begin{eqnarray}
S_B^\eta(p) \equiv - \eta_F(p) \left[ iS_B^V(p) - i\overline S_B^V(p) \right] \,,
\end{eqnarray}
and 
\begin{eqnarray}
\overline S_B^V(p) \equiv \gamma_0 S^{V \dagger}_B(p) \gamma_0 \,,
\label{Sbar}
\end{eqnarray}
for a fermion propagator, such that
\begin{eqnarray}
S_B^\eta(p) = - \eta_F(p) \int_{-\infty}^\infty ds\; e^{\Phi(p,s)} G(p,s) \,.
\label{Seta}
\end{eqnarray}
Here $\eta_F(p)$ contains the distribution function for the fermions and the anti-fermions:
\begin{eqnarray}
\eta_F(p) &=& \Theta(p\cdot u) f_F(p,\mu,\beta) \nonumber \\
&+& \Theta(-p\cdot u) f_F(-p,-\mu,\beta) \, ,
\label{eta}
\end{eqnarray}
 $f_F$ denotes the Fermi-Dirac distribution function:
\begin{eqnarray}
f_F(p,\mu,\beta) = {1\over e^{\beta(p\cdot u - \mu)} + 1} \,,
\label{distrib}
\end{eqnarray}
and $\Theta$ is the step function given by:
\begin{eqnarray}
\Theta(x) &=& 1, \; \mbox{for $x > 0$} \,, \nonumber \\
&=& 0, \; \mbox{for $x < 0$} \,. \nonumber                
\end{eqnarray}
Here the four velocity of the medium is $u$. In the rest frame its
components are $u^{\mu}=(1,0,0,0)$. 
%
\subsection{The expression for $\Pi^5_{\mu \nu}(k)$ in thermal medium and
in the presence of a background uniform magnetic field}
The relevant Feynman diagram of the process appears in
fig.~[\ref{f:cher}]. Following that diagram 
the axialvector-vector amplitude $\Pi^5_{\mu \nu}(k)$ is expressed as
\begin{eqnarray}
i\Pi^5_{\mu \nu}(k)=(-i e)^2(-1) \int {{d^4 p}\over {(2\pi)^4}}\Tr \left
[ \gamma_\mu \gamma_5 iS(p) \gamma_\nu iS(p')\right].
\label{pi5}
\end{eqnarray}
The vacuum part has already been done in {\cite{hari}} and the thermal
part with two factors of $\eta_F$ is related to pure absorption
effects in the medium, which we are leaving out for the time being.
The remaining terms are
\begin{eqnarray}
i\Pi^5_{\mu\nu}(k)&=& e^2 \int {{d^4 p}\over {(2\pi)^4}}\Tr \left
[ \gamma_\mu \gamma_5 S^\eta_B(p) \gamma_\nu
iS^V_B(p')\right.\nonumber\\
&+&\left. \gamma_\mu
\gamma_5 iS^V_B(p) \gamma_\nu S^\eta_B(p')\right].
\label{pi-ini}
\end{eqnarray}

Using the form of the fermion propagator in a magnetic field in
presence of a thermal medium, as given by expressions(\ref{SV}) and
(\ref{Seta}) we get
\begin{eqnarray}
i\Pi^5_{\mu\nu}(k)&=& -e^2 \int {{d^4 p}\over {(2\pi)^4}}
\int_{-\infty}^\infty ds\, e^{\Phi(p,s)}\nonumber\\
& &\int_0^\infty
ds'e^{\Phi(p',s')}\big[ \Tr \left[\gamma_\mu\gamma_5 G(p,s)
\gamma_\nu G(p',s')\right]\eta_F(p)\big. \nonumber\\
& &\big.\hskip 1cm + \Tr \left[\gamma_\mu
\gamma_5 G(-p',s') \gamma_\nu G(-p,s)\right]\eta_F(-p)
 \big]\nonumber\\
 &=& -e^2 \int {{d^4 p}\over {(2\pi)^4}}
\int_{-\infty}^\infty ds\, e^{\Phi(p,s)}\nonumber\\
& & \times  \int_0^\infty
ds'\,e^{\Phi(p',s')}\mbox{R}_{\mu\nu}(p,p',s,s')
\label{compl}
\end{eqnarray}
where $\mbox{R}_{\mu\nu}(p,p',s,s')$ contains the traces. 
\subsection{$\mbox{R}_{\mu\nu}$ to even and odd orders in magnetic field}
We calculate $\mbox{R}_{\mu\nu}(p,p',s,s')$ to even and odd orders in the
external magnetic field and call them $\mbox{R}^{(e)}_{\mu\nu}$
and $\mbox{R}^{(o)}_{\mu\nu}$. The reason for doing this is 
that the two contributions have different properties as far as their
dependence on medium is concerned, and the contributions are
\begin{eqnarray}
\mbox{R}^{(e)}_{\mu\nu}&=&
4i\eta_{-}(p)\varepsilon_{\mu \nu \alpha \beta}
\left[p^{\alpha\para} p'^{\beta\para}(1 + \tan(e{\mathcal B}s)\tan(e{\mathcal
B}s'))\right.\nonumber\\ 
&+&\left. p^{\alpha\para} p'^{\beta\pr} \sec^2 (e{\mathcal
B}s') 
+ p^{\alpha\pr} 
p'^{\beta\para} \sec^2 (e{\mathcal B}s)\right.\nonumber\\
&+&\left. p^{\alpha\pr} p'^{\beta\pr}\sec^2 (e{\mathcal
B}s)\sec^2 (e{\mathcal B}s')\right]
\label{reven}
\end{eqnarray}
and
\begin{eqnarray}
\mbox{R}^{(o)}_{\mu\nu}&=& 4i\eta_{+}(p)\left[ m^2\varepsilon_{\mu \nu
1 2}(\tan(e{\mathcal B}s) + \tan(e{\mathcal B}s'))\right.\nonumber\\
&+&\left.\left\{(g_{\mu \alpha\para} p^{\widetilde{\alpha\para}}
p'_{\nu\para} - g_{\mu \nu} p'_{\alpha\para}
p^{\widetilde{\alpha\para}} +g_{\nu \alpha\para} p^{\widetilde{\alpha\para}}
p'_{\mu\para} )\right.\right.\nonumber\\
&+&\left.\left.(g_{\mu \alpha\para} p^{\widetilde{\alpha\para}}
p'_{\nu\pr} + g_{\nu \alpha\para} p^{\widetilde{\alpha\para}}
p'_{\mu\pr}) \sec^2(e{\mathcal B}s')\right\} \tan(e{\cal
B}s)\right.\nonumber\\
&+& \left.\left\{(g_{\mu \alpha\para} p'^{\widetilde{\alpha\para}}
p_{\nu\para} - g_{\mu \nu} p_{\alpha\para}
p'^{\widetilde{\alpha\para}} +g_{\nu \alpha\para} 
p'^{\widetilde{\alpha\para}}
p_{\mu\para} )\right.\right.\nonumber\\
&+&\left.\left.(g_{\mu \alpha\para} p'^{\widetilde{\alpha\para}}
p_{\nu\pr} + g_{\nu \alpha\para} p'^{\widetilde{\alpha\para}}
p_{\mu\pr}) \sec^2(e{\mathcal B}s)\right\} \tan(e{\mathcal
B}s') \right]. 
\label{rodd}
\end{eqnarray}
Here
\begin{eqnarray}
\eta_+(p)&=&\eta_F(p) + \eta_F(-p) \label{etaplus}\\
\eta_-(p)&=&\eta_F(p) - \eta_F(-p)
\label{etaminus}
\end{eqnarray}
which contain the information about the distribution functions.
Also it should be noted that, in our convention
\begin{eqnarray}
a_{\mu} b^{{\widetilde \mu}\para}=a_0 b^3 + a_3 b^0.\nonumber
\end{eqnarray}
As stated we have split the contributions to $\Pi^5_{\mu\nu}(k)$ to
odd and even orders in the external constant magnetic field. The main
reason for doing so is the fact that, $\Pi^{5(o)}_{\mu\nu}(k)$ and
$\Pi^{5(e)}_{\mu\nu}(k)$, the axialvector-vector amplitude to odd and
even powers in $e{\mathcal B}$, have different dependence on the
background matter. Pieces proportional to even powers in ${\mathcal B}$
are proportional to $\eta_{-}(p_0)$, an odd function of the chemical
potential. On the other hand pieces proportional to odd powers in
${\mathcal B}$ depend on $\eta_{+}(p_0)$, and are even in $\mu$ and as a
result it survives in the limit $\mu \to 0$. This is a direct
consequence on the charge conjugation and parity symmetries of the
underlying theory.

From Eq.~(\ref{reven}) we notice that $\Pi^5_{\mu \nu}(k)$ to even orders
in the magnetic field satisfies the current conservation condition in
both the vertices. In Eq.~(\ref{rodd}) we see that all the terms in
the right hand side are symmetric in the $\mu$ and $\nu$ indices
except the first term. This term differentiates between the two
vertices in this case and as $\Pi^5_{\mu \nu}(k)$ to odd orders in
magnetic field is gauge invariant in the $\nu$ vertex we do not get
the same condition for the axialvector vertex. If in Eq.~(\ref{rodd})
we put $m=0$ then all the terms in the right will be symmetric in both
the tensor indices, and as a result current conservation condition
will hold for both vertices. If the mass of the looping fermion is not
zero then from the above analysis we can say that only Eq.~(\ref{ngi})
will hold. If the looping fermion is massless then Eq.~(\ref{mgi})
will hold, something which is expected.

If we concentrate on the rest frame of the medium, then $p\cdot
u=p_0$. Thus, the distribution function does not depend on the
spatial components of $p$.  From the form of Eq.~(\ref{reven})
and Eq.~(\ref{rodd}) we find that in
Eq.~(\ref{compl}), the integral over the transverse components of $p$
has the following generic structure:
\begin{eqnarray}
\int d^2 p_\perp \; e^{\Phi(p,s)} e^{\Phi(p',s')} \times
\mbox{($p^{\beta_\perp}$ or $p'^{\beta_\perp}$)} \,.
\end{eqnarray}
Notice now that
\begin{eqnarray}
{\partial \over \partial p_{\beta_\perp}} 
\Big[ \; e^{\Phi(p,s)} e^{\Phi(p',s')} \Big] = 
{2i\over e{\mathcal B}} \Big( \tan (e{\mathcal B}s) \; p^{\beta_\perp} + \tan
(e{\mathcal B}s') \; p'^{\beta_\perp} \Big)
e^{\Phi(p,s)} e^{\Phi(p',s')} \,.
\label{single_derivative}
\end{eqnarray}
However, this expression, being a total derivative, should integrate
to zero. Thus we obtain that 
\begin{eqnarray}
\tan (e{\mathcal B}s) \; p^{\beta_\perp} \stackrel\circ= - \tan
(e{\mathcal B}s') \; p'^{\beta_\perp} \,,
\end{eqnarray}
where the sign `$\stackrel\circ=$' means that the expressions on both
sides of it, though not necessarily equal algebraically, yield the
same integral. This gives
\begin{eqnarray}
p^{\beta_\perp} &\stackrel\circ=& - \, {\tan (e{\mathcal B}s') \over \tan
(e{\mathcal B}s) + \tan (e{\mathcal B}s')} \; k^{\beta_\perp}
\,,\label{pperpint}\nonumber\\* 
p'^{\beta_\perp} &\stackrel\circ=&  {\tan (e{\mathcal B}s) \over \tan
(e{\mathcal B}s) + \tan (e{\mathcal B}s')} \; k^{\beta_\perp} \,.
\label{primeperpint}
\end{eqnarray}
Similarly we can derive some other relations which can be used under
the momentum integral signs. To write them in a useful form, we turn
to Eq.\ (\ref{single_derivative}) and take another derivative with
respect to $p^{\alpha_\perp}$. From the fact that this derivative
should also vanish on $p$ integration, we find
\begin{eqnarray}
p_\perp^\alpha p_\perp^\beta \stackrel\circ= {1\over \tan (e{\cal
B}s) + \tan (e{\mathcal B}s')} \Bigg[{ie{\mathcal B} \over 2}
g_\perp^{\alpha\beta} +  {\tan^2 (e{\mathcal B}s') \over \tan (e{\cal
B}s) + \tan (e{\mathcal B}s')} \;
k_\perp^\alpha k_\perp^\beta \Bigg] \,.
\end{eqnarray}
In particular, then,
\begin{eqnarray}
p_\perp^2 \stackrel\circ= {1\over \tan (e{\cal
B}s) + \tan (e{\mathcal B}s')} \Bigg[ -ie{\mathcal B} +
{\tan^2 (e{\mathcal B}s') \over \tan (e{\mathcal B}s) + \tan (e{\mathcal B}s')} \;
k_\perp^2 \Bigg] \,.
\label{psq}
\end{eqnarray}
It then simply follows that
\begin{eqnarray}
p_\perp^{\prime2} \stackrel\circ= {1\over \tan (e{\cal
B}s) + \tan (e{\mathcal B}s')} \Bigg[ -ie{\mathcal B} +
{\tan^2 (e{\mathcal B}s) \over \tan (e{\mathcal B}s) + \tan (e{\mathcal B}s')} \;
k_\perp^2 \Bigg] \,.
\label{p'sq}
\end{eqnarray}
And finally using the definition of the exponential factor in
Eq.~(\ref{Phi}) we can erite
\begin{eqnarray}
m^2 &\stackrel{\circ}{=}&\left(i{d\over
ds} + (p^2\para -\sec^2(e{\mathcal B}s) p^2\pr)\right)\,.
\end{eqnarray}
Using the above relations we get
\begin{eqnarray}
\mbox{R}^{(e)}_{\mu \nu}&\stackrel{\circ}{=}&
4i\eta_{-}(p_0)\left[\varepsilon_{\mu \nu \alpha\para \beta\para}
p^{\alpha\para} p'^{\beta\para}(1 + \tan(e{\mathcal B}s)\tan(e{\mathcal B}s'))\right.\nonumber\\
&+&\left. \varepsilon_{\mu \nu \alpha\para
\beta\pr} p^{\alpha\para} p'^{\beta\pr} \sec^2 (e{\cal
B}s')\right.\nonumber\\
&+&\left.\varepsilon_{\mu \nu \alpha\pr \beta\para} p^{\alpha\pr} 
p'^{\beta\para} \sec^2 (e{\mathcal B}s)\right]
\label{reven1}
\end{eqnarray}
and 
\begin{eqnarray}
\mbox{R}^{(o)}_{\mu \nu}&\stackrel{\circ}{=}&
4i\eta_+(p_0)\left[-\varepsilon_{\mu \nu 1 2} 
\left\{ \frac{\sec^2(e{\mathcal B}s)\tan^2(e{\mathcal B}s')}{\tan(e{\mathcal B}s)
+ \tan(e{\mathcal B}s')}
k_{\pr}^2 \right.\right.\nonumber\\
&+& \left.\left. (k\cdot p)\para (\tan(e{\mathcal B}s) +
\tan(e{\mathcal B}s'))\right\}\right.\nonumber\\ 
&+&\left. 2\varepsilon_{\mu 1 2 \alpha\para}(p'_{\nu\para}
p^{\alpha\para}\tan(e{\mathcal B}s) +
p_{\nu\para}p'^{\alpha\para}\tan(e{\mathcal B}s'))\right.\nonumber\\
&+&\left. g_{\mu\alpha\para} k_{\nu\pr}\left\{p^{\widetilde
\alpha\para}(\tan(e{\mathcal B}s)
 - \tan(e{\mathcal B}s'))\right.\right.\nonumber\\
&-&\left.\left. k^{\widetilde \alpha\para}\,
{\sec^2(e{\mathcal B}s)\tan^2(e{\mathcal B}s')\over{\tan(e{\mathcal B}s) +
\tan(e{\mathcal B}s')}}\right\}\right.\nonumber\\
&+&\left.\{g_{\mu\nu}(p\cdot \widetilde k)\para + g_{\nu \alpha\para}
p^{\widetilde \alpha\para} k_{\mu\pr}\} \right.\nonumber\\
&\times& \left.(\tan(e{\mathcal B}s) - \tan(e{\mathcal B}s'))\right.\nonumber\\
&+&\left. g_{\nu \alpha\para}
k^{\widetilde\alpha\para}p_{\mu\pr}\sec^2(e{\mathcal B}s)\tan(e{\mathcal B}s')\right].
\label{crmunu}
\end{eqnarray}
Before going into the next section we comment on the nature of the
integral appearing in Eq.(\ref{compl}). The first point to make is
that from the form of $\mbox{R}^{(e)}_{\mu \nu}$ in Eq.~(\ref{reven1})
we note the axialvector-vector amplitude in a magnetized medium to even
orders in the magnetic field is antisymmetric, as it was in a medium
without any magnetic field. Contrary to this $\mbox{R}^{(o)}_{\mu
\nu}$ does not have any well defined symmetry property.

Secondly, as the integrals are not done explicitely something must be
said about the possible divergences which may creep up in evaluating
them. In principle we expect no divergences here. The reasons are as
follows. Firstly we are working in finite temperatures and so an
automatic ultraviolet cutoff, the temperature $T$ off the medium, is
already present. Secondly it must be noted that magnetic fields brings
in no new divergences in the calculations. The divergence that could
have been present would have come from the vacuum contribution of
$\Pi^5_{\mu \nu}(k)$ when ${\mathcal B} = 0$, but in this case that
part does not exist at all, as we have seen from Section
{\ref{gd}}. In this connection it can be said that in absence of the
medium but in presence of the background magnetic field another
divergent structure could arise, that is anomaly, due to the presence
of the axial vector vertex. Anomaly is essentialy an ultraviolet
phenomenon which shows up in non conservation of some currents, after
taking the quantum corrections, which were conserved classically. But
in the present case these need not worry us because we are working in
a thermal medium and as discussed previously the ultraviolet regulators
are already present in our theory.
 
As a result the integral expression for $\Pi^5_{\mu
\nu}(k)$ in our case does not have any singularities. So we can now
write the full expression of the axialvector-vector amplitude as
\begin{eqnarray}
i\Pi^5_{\mu\nu}(k)&=& -e^2 \int {{d^4 p}\over {(2\pi)^4}}
\int_{-\infty}^\infty ds\, e^{\Phi(p,s)}\nonumber\\
& & \times  \int_0^\infty
ds'\,e^{\Phi(p',s')}\left[\mbox{R}^{(o)}_{\mu\nu} + \mbox{R}^{(e)}_{\mu\nu}\right]
\label{alord}
\end{eqnarray}
where $\mbox{R}^{(o)}_{\mu\nu}$ and
$\mbox{R}^{(e)}_{\mu\nu}$ are given by
Eqs.(\ref{crmunu}) and (\ref{reven1}) in the rest frame of the
medium. 
\section{Zero momentum limit and effective charge}
\label{efftcharge}
The off-shell electromagnetic vertex function $\Gamma_{\nu}$ is
defined in such a way that, for on-shell neutrinos, the $\nu \nu
\gamma$ amplitude is given by:
\begin{eqnarray}
{\cal M} = - i \bar{u}(q') \Gamma_{\nu} u(q) A^{\nu}(k),
\label{chargedef}
\end{eqnarray}
where, $k$ is the photon momentum. Here, $u(q)$ is the the neutrino
spinor and $A^{\nu}$ stands for the electromagnetic vector
potential. In general $\Gamma_{\nu}$ would depend on $k$ and the
characteristics of the medium. With our effective Lagrangian in
Eq.~(\ref{effl}), $\Gamma_{\nu}$ is given by
\begin{eqnarray}
\Gamma_{\nu} = - \frac{1}{\sqrt{2}e} G_F \gamma^{\mu} (1 - \gamma_5) \,(g_{\rm V} \Pi_{\mu \nu} + g_{\rm A} \Pi_{\mu \nu}^5) \,,
\end{eqnarray}

The effective charge of the neutrinos is defined in terms of the
vertex function by the following relation~\cite{pal1}:
\begin{eqnarray}
e_{\rm eff} = {1\over{2 q_0}} \, \bar{u}(q) \, \Gamma_0(k_0=0, {\bf k}
\rightarrow 0) \, u(q) \,.
\label{chargedef1}
\end{eqnarray}
For massless Weyl spinors this definition can be rendered into the form:
\begin{eqnarray}
e_{\rm eff} = {1\over{2 q_0}} \, \Tr  \left[\Gamma_0(k_0=0, {\bf
k}\rightarrow 0) \, (1+\lambda \gamma^5) \, \rlap/q \right]
\label{nec1}
\end{eqnarray}
where $\lambda = \pm 1$ is the helicity of the spinors.

We have remarked earlier in Section \ref{gd}, that in a medium, we
have an additional vector $u^{\mu}$.  The axialvector-vector amplitude
in this case, of the form $\varepsilon_{\mu \nu \alpha
\beta}u^{\alpha}k^{\beta}$, do not contribute for the effective
electric charge of the neutrinos since for charge calculation we have
to put the index $\nu = 0$. In the rest frame only the time component
of the four vector $u$ exists, that forces the totally antisymmetric
tensor to vanish. But the polarization tensor can be expanded in terms
of form factors along with the new tensors constructed out of
$u^{\mu}$ and the ones we already had in absence of a medium as,
\begin{eqnarray}
\Pi_{\mu\nu}(k)=\Pi_T\,T_{\mu\nu} + \Pi_L\,L_{\mu\nu}. 
\end{eqnarray}
Here
\begin{eqnarray}
T_{\mu\nu}&=& {\widetilde g}_{\mu\nu} - L_{\mu\nu}\nonumber\\
L_{\mu\nu}&=&\frac{{\widetilde u}_{\mu}{\widetilde
u}_{\nu}}{{\widetilde u}^2}\nonumber
\label{t}
\end{eqnarray}
with
\begin{eqnarray}
{\widetilde g}_{\mu\nu}&=&g_{\mu\nu} - \frac{k_{\mu}k_{\nu}}{k^2}\nonumber\\
{\widetilde u}_{\mu}&=&{\widetilde g}_{\mu\rho} u^{\rho}\nonumber
\end{eqnarray}
The longitudinal projector $L_{\mu \nu}$ is not zero in the limit
$k_0=0,\vec{k}\rightarrow 0$ and $\Pi_L$ is also not zero in the above
mentioned limit\cite{palone}. This fact is responsible for giving
nonzero contribution to the effective charge of neutrino in a medium.

From the Eq.~(\ref{harid}) we see that the axialvector-vector
amplitude in a background magnetic field without any medium does not
survive when the momentum of the external photon vanishes, and as a
result there cannot be any effective electric charge of the neutrinos
in a constant background magnetic field.  Actually this formal
statement could have been spoilt by the presence of possible infrared
divergence in the loop; i.e to say in $C\para$ and $C\pr$
\cite{iorf}. Since the particle inside the loop is massive so there is
no scope of having infrared divergence, hence it doesn't contribute to
neutrino effective charge.

Now we concentrate on the zero momentum limit of that part of the axial
polarization tensor which is going to contribute for the neutrino
effective charge in a magnetized plasma. From the onset it is to be
made clear that we are only calculating the axial contribution to the
effective charge.
\subsection{Effective charge to odd orders in external field}
Denoting $\Pi^{5}_{\mu\nu}(k_0=0, {\vec k } \to 0)$ by
$\Pi^{5}_{\mu\nu}$, we obtain
\begin{eqnarray}
\Pi^5_{\mu 0}&=&\lim_{k_0=0
\vec{k}\rightarrow 0}4 e^2 \int{d^4p\over{(2\pi)^4}} 
\int^{\infty}_{-\infty} ds\, e^{\Phi(p,s)}
\nonumber \\ 
&\times&\int^{\infty}_0 ds'
e^{\Phi(p',s')}(\tan(e{\mathcal B}s) + \tan(e{\mathcal B}s')) \nonumber\\
&\times &\eta_+(p_0)\left[2 p^2_0 -  (k\cdot p)\para \right]\varepsilon_{\mu 0 1 2}
\label{pi5k0}
\end{eqnarray}
the other terms turns out to be zero in this limit. The above equation
shows that, except the exponential functions, the integrand is free of
the perpendicular components of momenta. This is a peculiarity of this
case that the perpendicular excitations of the loop momenta are only
present in the phase like part of the integrals and in effect
decouples from the seen once they are integrated out. Its presence is
felt only through a linear dependence of the external field ${\mathcal
B}$ when the perpendicular components of $k$ vanish. Upon performing
the gaussian integration over the perpendicular components and taking
the limit $k_{\pr} \to 0$, we obtain,
\begin{eqnarray}
\Pi^5_{3 0}&=&\lim_{k_0=0,\vec{k}\rightarrow 0} -\frac{(-4i e^3
{\mathcal B})}{4\pi} \int{d^2 p\para \over{(2\pi)^2}} 
\int^{\infty}_{-\infty} ds \nonumber \\
&\times&
\, e^{is(p^2\para - m^2) - \varepsilon|s|}
\int^{\infty}_0 ds'
e^{is'(p'^{\,2}\para - m^2) - \varepsilon|s'|} \nonumber\\
&\times &\eta_+(p_0)\left[2 p^2_0 -  (k\cdot p)\para
\right].
\label{charge1}
\end{eqnarray}
It is worth noting that the $s$ integral gives
\begin{eqnarray}
\int^{\infty}_{\infty} ds\, e^{is(p^2\para - m^2) - \varepsilon|s|} =
2\pi \delta(p^2\para - m^2)
\label{delta}
\end{eqnarray}
and the $s'$ integral gives
\begin{eqnarray}
\int^{\infty}_0 ds'\, e^{is'(p'^{\,2}\para - m^2) - \varepsilon|s'|} =
{i\over{(p'^{\,2}\para - m^2) + i\varepsilon}}.
\label{divergent}
\end{eqnarray}
Using the above results in Eq.(\ref{charge1}) and using the delta
function constraint, we arrive at,
\begin{eqnarray}
\Pi^5_{3 0} &=&
\lim_{k_0=0,\vec{k}\rightarrow 0} 
- 2(e^3 {\mathcal B}) \int{d^2 p\para \over{(2\pi)^2}}
{\delta(p^2\para - m^2)}
\eta_+(p_0)
\nonumber \\ &\times&
\Bigg[{2 p^2_0\over{(k^{\,2}\para +2(p.k)\para)}}
 -  \frac{1}{2}\Bigg].
\label{charge2}
\end{eqnarray}
In deriving Eq.~(\ref{charge2}), pieces proportional to $k^2\para$ in
the numerator were neglected. Now if one makes the substitution, $
p'\para \to (p\para + k\para/2) $ and sets $k_0 =0$ one arrives at,
\begin{eqnarray}
\Pi^5_{3 0} 
&=&
\lim_{k_0=0,\vec{k}\rightarrow 0} 
2(e^3 {\mathcal B}) \int{d p_3 \over{(2\pi)^2}}
\left(n_+(E'_p)+ n_-(E'_p)\right)
\nonumber \\ &\times&
\Bigg[ \frac{E'_p}{p_3k_3}
 + \frac{1}{2E'_p}\Bigg].
\label{charge3}                
\end{eqnarray}
Here  $n_{\pm}(E'_p)$ are the functions $f_F(E'_p,-\mu,\beta)$,
and $f_F(E'_p,\mu,\beta)$, as given in Eq.(\ref{distrib}), which are
nothing but the Fermi-Dirac distribution functions of the particles
and the antiparticles in the medium with a modified energy $E'_p$.
The new term $E'_p$  is defined as follows,
\begin{eqnarray}
E'^2_p=[(p_3-k_3/2)]^2 + m^2, \nonumber
\end{eqnarray}
and it can be expanded for small external
momenta in the following way
\begin{eqnarray}
 E'^2_p\simeq p^2_3+m^2 - p_3 k_3 = E^2_p - p_3 k_3\nonumber
\end{eqnarray}
where $E^2_p = p^2_3+m^2$.
Noting, that
\begin{eqnarray}
E'_p= E_p -  {p_3 k_3 \over 2 E_p} + O(k^2_3),
\label{epprime}
\end{eqnarray}
one can use this expansion in Eq.~(\ref{charge3}), to arrive at:
\begin{eqnarray}
\Pi^5_{3 0} 
&=&
\lim_{k_0=0,\vec{k}\rightarrow 0} 
2(e^3 {\mathcal B}) \int{d p_3 \over{(2\pi)^2}}
\left(n_+(E'_p)+ n_-(E'_p)\right) 
\Bigg[ \frac{E_p}{p_3 k_3}
 \Bigg].
\label{charge5}
\end{eqnarray}
The expression for for $\eta_+(E'_p) = n_+(E'_p)+ n_-(E'_p)$ when
expanded in powers of the external momentum $k_3$ is given by
\begin{eqnarray}
\eta_+(E'_p) = ( 1 + \frac{1}{2} \frac{\beta p_3 k_3}{E_p}) \eta_+(E_p)
\label{neweta}
\end{eqnarray}
up to first order terms in the external momentum $k_3$.

\subsubsection{Effective charge For $\mu \ll m$}

In the limit, when $\mu \ll m$ one can use the 
following expansion,
\begin{eqnarray}
\eta_+(E'_p) &=&\Bigg[ n_{+}(E'_p) +n_{-}(E'_P)\Bigg]\nonumber\\
&=& 2 \sum_{n=0}^{\infty} (-1)^n \cosh([n+1]\beta\mu)e^{-(n+1)\beta E_p}
\nonumber \\ &\times&
\left(1 +\frac{\beta p_3 k_3}{2E_p} + O(k_3^2)+ .....\right)
\label{expan}
\end{eqnarray}

Now  using Eq.(\ref{expan}) in Eq.(\ref{charge5}) we get
\begin{eqnarray}
\Pi^5_{3 0} &=&
\lim_{k_0=0, \vec{k}\rightarrow 0} 
(4e^3 {\mathcal B}) 
\sum_{n=0}^{\infty} (-1)^n\nonumber\\
 &\times& \cosh([n+1]\beta\mu)
\int{d p_3 \over{(2\pi)^2}}e^{-(n+1)\beta E_p}
\nonumber \\
 &\times& \Bigg[ \frac{E_p}{(p_3 k_3)} + \frac{\beta}{2}
 \Bigg].
\label{charge6}
\end{eqnarray}
The first term vanishes by symmetry of the integral, but the second term 
is finite and so we get:
\begin{eqnarray}
\Pi^5_{3 0} &=& \beta 
\lim_{k_0=0 {\vec k}\to 0}
\frac{(e^3 {\mathcal B})}{2\pi^2} 
\sum_{n=0}^{\infty} (-1)^n \cosh([n+1]\beta\mu)
\nonumber
\\ &\times&
\int d p_3  e^{-(n+1)\beta E_p}.
\label{charge7}
\end{eqnarray} 

To perform the momentum integration, use of the following integral
transform turns out to be extremely convenient
\begin{eqnarray}
e^{-\alpha \sqrt{s}} =\frac{\alpha}{2\sqrt{\pi}} 
\int^{\infty}_0 du e^{-us - \frac{\alpha^2}{4u}} u^{-3/2}.  
\end{eqnarray}
Identifying $\sqrt{s}$ with $E_p$ and $[(n+1)\beta]$ as $\alpha$ one
can easily perform the gaussian $p_3$ integration without any
difficulty. The result is:
\begin{eqnarray}
\Pi^{5}_{3 0}&=& \beta
\frac{(e^3 {\mathcal B})}{2\pi^2} 
\sum_{n=0}^{\infty} (-1)^n \cosh([n+1]\beta\mu)
\nonumber
\\ &\times&
 (\beta(n+1)/2)\int du e^{- m^2 u - \frac{((n+1)\beta/2)^2}{u}} u^{-2}.
\label{charge8}
\end{eqnarray}
Performing the the u integration
the axial part of the effective charge of neutrino in the limit of
$m > \mu $ turns out to be,
\begin{eqnarray}
e^{\nu_a}_{\rm eff} &=& -  \sqrt{2} g_{A} m \beta G_F
\frac{e^2 {\mathcal B}}{\pi^2} (1-\lambda) \cos(\theta)\nonumber\\
&\times&\sum_{n=0}^{\infty} (-1)^n \cosh((n+1)\beta\mu)  
 K_{-1}(m \beta(n+1)).
\label{nucharge}
\end{eqnarray}
Here $\theta$ is the angle between the neutrino three momentum and the
background magnetic field. The superscript $\nu_a$ on $e^{\nu_a}_{\rm
eff}$ denotes that we are calculating the axial contribution of the
effective charge. $K_{-1}(m \beta(n+1))$ is the modified Bessel
function (of the second kind) of order one (for this function
$K_{-1}(x) = K_{1}(x)$) which sharply falls off as we move away from
the origin in the positive direction.  Although as temperature tends
to zero Eq.~(\ref{nucharge}) seems to blow up because of the presence
of $m \beta$, but $K_{-1}(m\beta(n+1))$ would damp it's growth as
$e^{-m\beta}$, hence the result remains finite.
\subsubsection{Effective charge for $\mu \gg m$}
Here we would try to estimate neutrino effective charge when $\mu \gg
m$. Using Eqs.(\ref{charge5}) and (\ref{neweta}) we would obtain
\begin{eqnarray}
\Pi^5_{3 0} = \frac{e^3 {\mathcal B}}{
2\pi} \beta \int \frac{dp}{2\pi} \eta_+(E_p).
\label{one}
\end{eqnarray}
Neglecting  $m$ in the expression in $E_p$ we would obtain, 
\begin{eqnarray}
\Pi^5_{3 0} = \frac{e^3 {\mathcal B}}{
2\pi^2} \ln[(1 +  e^{\beta \mu})( 1 +  e^{-\beta \mu})].
\end{eqnarray}
Same can also be written as
\begin{eqnarray}
\Pi^5_{3 0} 
=\frac{e^3 {\mathcal B}}{\pi^2} \ln \left(2\cosh(\frac{\beta \mu}{2})\right).
\label{ra}
\end{eqnarray}
The expression for the effective charge then turns out to be
\begin{eqnarray}
e^{\nu_a}_{\rm eff} = - \sqrt{2} g_A G_F \frac{e^2 {\mathcal B}}{\pi^2} \ln
\left(2\cosh(\frac{\beta \mu}{2})\right) (1- \lambda) \cos(\theta)
\end{eqnarray}
where $\lambda$ is the helicity of the neutrino spinors.

Before going to the next section some general discussion about the
effective charge expression can be made. In a background magnetic
field the field dependence of the form factors, which are usually
scalars, can be of the following form:
\begin{eqnarray}
k^{\mu}F_{\mu\nu}F^{\nu\lambda}k_{\lambda} 
\mbox{~~~and~~~}
F_{\mu\nu}F^{\mu\nu}\,,  
\end{eqnarray}
or
\begin{eqnarray}
({\widetilde F}u)^{\mu}({\widetilde F}u)_{\mu}\,.
\end{eqnarray}
These forms dosen't exhaust all the possibilities, other terms can
also be constructed by the above forms. The thing which must be
noted is when $k$ tends to zero only terms that can survive in the
form factors must be an even function of ${\mathcal B}$.

Of all possible tensorial structures for the axialvector-vector amplitude
in a magnetized plasma, there exists one which satisfies the current
conservation condition in the $\nu$ vertex and is given by,
\begin{eqnarray}
{\widetilde F}_{\mu \alpha} u^{\alpha} 
u^{\para}_{\nu}\,.\nonumber  
\label{ww}
\end{eqnarray}
Its worth noting this term in
Eq.~(\ref{ww}), which is odd in the external field, survives in the
zero external momentum limit in the rest frame of the medium. We have
earlier noted that the form factors which exist in the rest frame of
the medium and in the zero momentum limit are even in powers of the
external field. This tells us directly that the axial polarization
tensor must be odd in the external field in the zero external momentum
limit, a result which we have verified in this work.
\section{Conclusion}
\label{concl}
In this work we have elucidated upon the physical significance of the
axialvector-vector amplitude in various neutrino mediated processes in a
magnetized medium. We have analysed its gauge invariance
properties. Its tensor structure has been written down, and we have
shown the integral expression of the tensor is ultraviolet finite.  It
has been shown that the part of $\Pi^5_{\mu \nu}(k)$ even in $\mathcal B$
doesn't contribute to the effective electric charge. However it does
contribute to physical proceses e.g., neutrino Cherenkov radiation or
neutrino decay in a medium. It is worth noting that in the low density
high temperature limit, the magnitude of $e^{\nu_a}_{\rm eff}$ can
become greater than the effective charge of the neutrino in ordinary
medium provided $e{\mathcal B}$ is large enough. On the other hand in the
high density limit $e^{\nu_a}_{\rm eff}$ can dominate over the
effective charge of the neutrino as found in an unmagnetized medium,
provided the temperature is low enough. However in standard
astrophysical objects, e.g., core of type II Supernova temperature is
of the order of 30- 60 MeV with Fermi momentum around 300 MeV, for red
giants the same are 10 keV and 400 keV, for young white dwarves
temperature is around 0.1 - 1 keV and Fermi momentum 500 keV. In these
systems one can have relatively large induced neutrino charge,
provided the field strength is large enough.
\section*{Acknowledgment}
We would like to thank Prof. Palash B. Pal and Prof. Parthasarathi Majumdar
for helpful discussions comments and suggestions.
\appendix\section*{\hfil Appendix \hfil}
\section{Gauge invarience}
\label{gaugeinvariance}
Now we concentrate on Eq.~(\ref{ngi}), about which we discussed in
Section \ref{form}. The axialvector-vector amplitude has an
electromagnetic vertex and as a result electromagnetic current must be
conserved. From Eq.~(\ref{harid}) we see that $\Pi^{5}_{\mu\nu}(k)$ is
gauge invarient in the $\mu$-vertex, which is the electromagnetic
vertex in that case. In our case as discussed the $\nu$-vertex is the
electromagnetic vertex, and we explicitly show the gauge invariance
in that vertex below.
\subsection{Gauge invariance for $\Pi^5_{\mu \nu}(k)$ to even
orders in the external field}
The  axialvector-vector amplitude even in the external field is given by
\begin{eqnarray}
\Pi^{5(e)}_{\mu \nu}(k)&=& -(-i e)^2(-1) \int {{d^4 p}\over {(2\pi)^4}}
\int_{-\infty}^\infty ds\, e^{\Phi(p,s)}\nonumber\\
& & \times  \int_0^\infty
ds'\,e^{\Phi(p',s')}\mbox{R}^{(e)}_{\mu\nu}(p,p',s,s').
\label{a1}
\end{eqnarray}
Noting that, 
\begin{eqnarray}
q^{\alpha}p_{\alpha} = q^{\alpha\para}p_{\alpha\para} + 
q^{\alpha\pr}p_{\alpha\pr}\nonumber
\end{eqnarray}
we can write Eq.(\ref{reven1}) as,  
\begin{eqnarray}
\mbox{R}^{(e)}_{\mu \nu}&\stackrel{\circ}{=}&
4i\eta_{-}(p_0)\left[(\varepsilon_{\mu \nu \alpha \beta}
p^{\alpha} p'^{\beta} - \varepsilon_{\mu \nu \alpha \beta\pr}
p^{\alpha} p'^{\beta\pr}\right.\nonumber\\
&-&\left.\varepsilon_{\mu \nu \alpha\pr \beta}
p^{\alpha\pr} p'^{\beta})(1 + \tan(e{\mathcal B}s)\tan(e{\mathcal B}s'))\right.\nonumber\\
&+&\left. \varepsilon_{\mu \nu \alpha
\beta\pr} p^{\alpha} p'^{\beta\pr} \sec^2 (e{\cal
B}s')
+\varepsilon_{\mu \nu \alpha\pr \beta} p^{\alpha\pr} 
p'^{\beta} \sec^2 (e{\mathcal B}s)\right].\nonumber\\
\label{a3}
\end{eqnarray}
Here throughout we have omitted terms such as $\varepsilon_{\mu \nu
\alpha\pr \beta\pr} p^{\alpha\pr}  
p'^{\beta\pr}$, since  by the application of Eq.(\ref{pperpint}) we
have
\begin{eqnarray}
\varepsilon_{\mu \nu \alpha\pr \beta\pr} p^{\alpha\pr} 
p'^{\beta\pr} &=&\varepsilon_{\mu \nu \alpha\pr \beta\pr}
p^{\alpha\pr} p^{\beta\pr} + \varepsilon_{\mu \nu \alpha\pr
\beta\pr} p^{\alpha\pr}  
k^{\beta\pr} \nonumber\\
&\stackrel{\circ}{=}& - \frac{\tan(e{\mathcal B}s')}{\tan(e{\mathcal B}s')+\tan(e{\mathcal B}s')}
\varepsilon_{\mu \nu \alpha\pr 
\beta\pr} k^{\alpha\pr}  
k^{\beta\pr}\nonumber
\end{eqnarray}
which is zero.

After rearranging the terms appearing in Eq.(\ref{a3}), and by the
application of Eqs.(\ref{pperpint}) and (\ref{primeperpint}) we
arrive at the expression
\begin{eqnarray}
\mbox{R}^{(e)}_{\mu
\nu}&\stackrel{\circ}{=}&4i\eta_{-}(p_0)\Bigg[\varepsilon_{\mu \nu 
\alpha \beta} 
p^{\alpha} k^{\beta}(1 + \tan(e{\mathcal B}s)\tan(e{\mathcal B}s'))\nonumber\\
&+& \varepsilon_{\mu \nu \alpha
\beta\pr} k^{\alpha} k^{\beta\pr} 
\tan(e{\mathcal B}s) \tan(e{\mathcal B}s') \frac{\tan(e{\mathcal B}s)-\tan(e{\cal
B}s')}{\tan(e{\mathcal B}s) + \tan(e{\mathcal B}s')}\Bigg].\nonumber\\ 
\label{evenpart}
\end{eqnarray}
Because of the presence of terms like $\varepsilon_{\mu \nu
\alpha \beta} 
 k^{\beta}$ and $ \varepsilon_{\mu \nu \alpha
\beta\pr} k^{\alpha} $ if we contract $\mbox{R}^{(e)}_{\mu \nu}$  by
$k^\nu$, it vanishes.
\subsection{Gauge invariance for $\Pi^5_{\mu \nu}(k)$ to odd
orders in the external field}
The  axialvector-vector amplitude odd in the external field is given by
\begin{eqnarray}
\Pi^{5(o)}_{\mu \nu}(k)&=& -(-i e)^2(-1) \int {{d^4 p}\over {(2\pi)^4}}
\int_{-\infty}^\infty ds\, e^{\Phi(p,s)}\nonumber\\
& & \times  \int_0^\infty
ds'\,e^{\Phi(p',s')}\mbox{R}^{(o)}_{\mu\nu}(p,p',s,s')
\end{eqnarray}
where $\mbox{R}^{(o)}_{\mu\nu}(p,p',s,s')$ is given by Eq.(\ref{crmunu}).
The general gauge invariance condition in this case
\begin{eqnarray}
k^{\nu} \Pi^{5(o)}_{\mu \nu}(k) &=& 0 
\end{eqnarray}
can always be written down in terms of the following two equations,
\begin{eqnarray}
k^{\nu} \Pi^{5(o)}_{\mu\para \nu}(k) &=& 0 \label{gipara}\\
k^{\nu} \Pi^{5(o)}_{\mu\pr \nu}(k) &=& 0
\label{gipr}
\end{eqnarray}
where $ \Pi^{5(o)}_{\mu\para \nu}(k)$  is that part of $
\Pi^{5(o)}_{\mu \nu}(k)$ where the index $\mu$ can take the values $0$ and 
$3$ only. Similarly 
 $ \Pi^{5(o)}_{\mu\pr \nu}(k)$ stands for the part of $
\Pi^{5(o)}_{\mu \nu}(k)$ where  $\mu$  can take the values $1$ and $2$
only.  $ \Pi^{5(o)}_{\mu\para \nu}(k)$ contains $\mbox{R}^{(o)}_{\mu\para
\nu}(p,p',s,s')$ which from Eq.(\ref{crmunu}) is as follows,
\begin{eqnarray}
\mbox{R}^{(o)}_{\mu\para \nu}&\stackrel{\circ}{=}&
4i\eta_+(p_0)
\left[
-\varepsilon_{\mu\para \nu 1 2} 
\left\{ \frac{\sec^2(e{\mathcal B}s)\tan^2(e{\mathcal B}s')}{\tan(e{\mathcal B}s)
+ \tan(e{\mathcal B}s')} 
k_{\pr}^2 \right.\right.\nonumber\\
&+& \left.\left. (k\cdot p)\para (\tan(e{\mathcal B}s) +
\tan(e{\mathcal B}s'))\right\}\right.\nonumber\\ 
&+&\left. 2\varepsilon_{\mu\para 1 2 \alpha\para}\,(p'_{\nu\para}
p^{\alpha\para}\tan(e{\mathcal B}s) +
p_{\nu\para}p'^{\alpha\para}\tan(e{\mathcal B}s'))\right.\nonumber\\
&+&\left. g_{\mu\para \alpha\para} k_{\nu\pr}\left\{p^{\widetilde
\alpha\para}(\tan(e{\mathcal B}s)
 - \tan(e{\mathcal B}s'))\right.\right.\nonumber\\
&-&\left.\left. k^{\widetilde \alpha\para}\,
{\sec^2(e{\mathcal B}s)\tan^2(e{\mathcal B}s')\over{\tan(e{\mathcal B}s) +
\tan(e{\mathcal B}s')}}\right\}\right.\nonumber\\
&+&\left. g_{\mu\para\nu}(p\cdot \widetilde k)\para(\tan(e{\mathcal B}s) -
\tan(e{\mathcal B} s'))
\right]
\label{para}
\end{eqnarray}
and $ \Pi^{5(o)}_{\mu\pr \nu}$ contains
$\mbox{R}^{(o)}_{\mu\pr \nu}(p,p',s,s')$  which is 
\begin{eqnarray}
{\mbox R}^{(o)}_{\mu\pr
\nu}&\stackrel{\circ}{=}&4i\eta_+(p_0)\left[\{g_{\mu\pr \nu}(p\cdot
\widetilde k)\para + g_{\nu \alpha\para} p^{\widetilde \alpha\para}
k_{\mu\pr}\}\right.\nonumber\\
& &\hskip 1cm \times  \left.(\tan(e{\mathcal B}s) - \tan(e{\mathcal B}s'))\right.\nonumber\\ 
&+& \left.g_{\nu \alpha\para}
k^{\widetilde\alpha\para}p_{\mu\pr}\sec^2(e{\mathcal B}s)\tan(e{\mathcal B}s')\right]. 
\label{perp}
\end{eqnarray}
Eqs.(\ref{gipara}),(\ref{gipr})  implies one should have the
following relations satisfied,
\begin{eqnarray}
k^{\nu} \int {{d^4 p}\over {(2\pi)^4}} 
\int_{-\infty}^\infty ds\, e^{\Phi(p,s)} \int_0^\infty
ds'\,e^{\Phi(p',s')} \, 
{\mbox R}^{(o)}_{\mu\pr \nu}=0
\nonumber \\
\label{gipr1}
\end{eqnarray}
and
\begin{eqnarray}
k^{\nu}\int {{d^4 p}\over {(2\pi)^4}} 
\int_{-\infty}^\infty ds\, e^{\Phi(p,s)} \int_0^\infty
ds'\,e^{\Phi(p',s')}\, {\mbox R}^{(o)}_{\mu\para \nu}=0.
\label{gipara1}
\end{eqnarray}
Out of the two above equations,  Eq.(\ref{gipr1}) can be verified 
easily since
\begin{eqnarray}
k^{\nu}{\mbox R}_{\mu\pr \nu}=0.
\end{eqnarray}

Now we look at Eq.(\ref{gipara1}). 
We explicitly consider the case $\mu_{\parallel}=3$ (the
$\mu_{\parallel}=0$ case lead to similar result). For
$\mu_{\parallel}=3$
\begin{eqnarray}
k^{\nu}{\mbox R}^{(o)}_{3 \nu}&\stackrel{\circ}{=}&-p_0\left[(p'^{\,2}\para -
p^2\para)(\tan(e{\mathcal B}s) + \tan(e{\mathcal B}s')) \right.\nonumber\\
&-&\left. k^2\pr(\tan(e{\mathcal B}s) - \tan(e{\mathcal B}s'))\right](4i \eta_+(p_0)).\nonumber\\
\label{a_1}
\end{eqnarray}

Apart from the small convergence factors, 
\begin{eqnarray}
{i \over e{\mathcal B}} && \left(\Phi(p,s) + \Phi(p',s')\right) \nonumber \\
&=& \left( p_\parallel^{\prime2} + p_\parallel^2 - 2m^2 \right) \xi 
- \left(p_\parallel^{\prime2} - p_\parallel^2 \right) \zeta \nonumber \\
&& - p\pr^{\prime2} \tan (\xi-\zeta) - p\pr^2 \tan (\xi+\zeta) \,,
\end{eqnarray}
\label{newa}
where we have defined the parameters
\begin{eqnarray}
\xi &=& \frac12 e{\mathcal B}(s+s') \,, \nonumber\\*
\zeta &=& \frac12 e{\mathcal B}(s-s') \,.
\label{xizeta}
\end{eqnarray}
From the last two equations we can write
\begin{eqnarray}
{ie{\mathcal B}} \; {d\over d\zeta} e^{\Phi(p,s) + \Phi(p',s')} 
&=& e^{\Phi(p,s) + \Phi(p',s')} \nonumber \\
& \times &\, \left(p_\parallel^{\prime2} - p_\parallel^2 - p\pr^{\prime2} \sec^2 (\xi-\zeta) + p\pr^2 \sec^2 (\xi+\zeta) \right) \,
\label{C1par}
\end{eqnarray}
which implies
\begin{eqnarray}
p'^{\,2}\para - p^2\para &=&
ie{\mathcal B}{d\over{d\xi}} + 
\left[ p'^{\,2}\pr
\sec^2(e{\mathcal B}s') - p^2\pr \sec^2(e{\mathcal B}s)\right].       
\label{a_parsqrdiff}
\end{eqnarray}
The equation above is valid in the sense that both sides of it
actually acts upon $e^{\widetilde \Phi(p,s,p',s')}$, where
\begin{eqnarray}
\widetilde \Phi(p,p',s,s') = \Phi(p,s) + \Phi(p',s').
\label{a_Phi}
\end{eqnarray}
From Eqs.(\ref{a_1}) and (\ref{a_parsqrdiff}) we have
\begin{eqnarray}
k^{\nu} \, {\mbox R}_{3 \,\nu}\, e^{\widetilde \Phi}
&\stackrel{\circ}{=}&-4i\eta_+(p_0)p_0\left[(p'^{\,2}\pr \sec^2(e{\mathcal B}s) - p^2\pr
\sec^2(e{\mathcal B}s))\right.\nonumber\\
& & \hskip 1cm \left.\times (\tan(e{\mathcal B}s)+\tan(e{\mathcal B}s'))\right.\nonumber\\
&-&\left. k^2\pr(\tan(e{\mathcal B}s)-\tan(e{\mathcal B}s'))\right.\nonumber\\
&+&\left.ie{\mathcal B}p_0(\tan(e{\mathcal B}s) + \tan(e{\mathcal B}s'))\,{d
\over{d \xi}}\right] e^{\widetilde 
\Phi}.\nonumber\\
\label{a_kdotR1}
\end{eqnarray}
Now using the the expressions for $p^2\pr$ and $p'^2\pr$ from
Eqs.(\ref{psq}) and (\ref{p'sq}) 
we can write
\begin{eqnarray}
k^{\nu} \, {\mbox R}_{3 \,\nu}\,e^{\widetilde \Phi}&\stackrel{\circ}{=}&4e{\mathcal B}\eta_+(p_0)
p_0\left[(\sec^2(e{\mathcal B}s) - \sec^2(e{\mathcal B}s'))\right.\nonumber\\
&+&\left. (\tan(e{\mathcal B}s) + \tan(e{\mathcal B}s')){d
\over{d \xi}}\right] e^{\widetilde \Phi} .\nonumber\\
\label{a_kdotR2}
\end{eqnarray}
The above equation can also be written as 
\begin{eqnarray}
k^{\nu} \, {\mbox R}_{3 \,\nu}\, e^{\widetilde
\Phi}\stackrel{\circ}{=}4e{\mathcal B} \eta_+(p_0)p_0 
{d \over{d \xi}} \left[ e^{\widetilde \Phi} (\tan(e{\mathcal B}s) +
\tan(e{\mathcal B}s'))\right].
\end{eqnarray}
Transforming to $\xi, \zeta$ variables and using the above
equation we can write the parametric integrations (integrations over
$s$ and $s'$) on the left hand side of Eq.(\ref{gipara1}) as
\begin{eqnarray}
\int_{-\infty}^{\infty} ds \int_0^{\infty} ds' k^{\nu} \, {\mbox R}_{3\,
\nu}\, e^{\widetilde \Phi}
=\frac{8\eta_+(p_0)p_0}{e{\mathcal B}}\int_{-\infty}^{\infty} d\xi
\int_{-\infty}^{\infty} d\zeta \Theta 
(\xi - \zeta) {d\over{d\xi}}{\cal F}(\xi,\zeta)\nonumber
\label{odd}
\end{eqnarray}
where
\begin{eqnarray}
{\cal F}(\xi,\zeta)=e^{\widetilde \Phi} (\tan(e{\mathcal B}s) +
\tan(e{\mathcal B}s')).\nonumber
\end{eqnarray}
The integration over the $\xi$ and $\zeta$ variables in Eq.(\ref{odd}) 
can be represented as,
\begin{eqnarray}
& &\int_{-\infty}^{\infty} d\xi \int_{-\infty}^{\infty} d\zeta \Theta
(\xi - \zeta) {d\over{d\xi}}{\cal
F}(\xi,\zeta)\nonumber\\
&=&\int_{-\infty}^{\infty} d\xi \int_{-\infty}^{\infty}
d\zeta \left[{d\over{d\xi}}\{ \Theta(\xi - \zeta) {\cal F}(\xi,\zeta)\}
- \delta(\xi - \zeta) {\cal F}(\xi,\zeta)\right]\nonumber\\
&=&-\int_{-\infty}^{\infty} d\xi {\cal F}(\xi,\xi)
\label{end1}
\end{eqnarray}
here the second step follows from the first one as the first integrand
containing the $\Theta$ function vanishes at both limits of the
integration. The remaining integral is now only a function of $\xi$
and is even in $p_0$. But in Eq.~(\ref{odd}) we have $\eta_+(p_0)p_0$
sitting, which makes the the integrand odd under $p_0$ integration in
the left hand side of Eq.~(\ref{gipara1}), as $\eta_+(p_0)$ is an even
function in $p_0$. So the $p_0$ integral as it occurs in the left hand
side of Eq.~(\ref{gipara1}) vanishes as expected, yeilding the
required result shown in Eq.~(\ref{gipara}).

\end{document}